\documentclass[%
reprint,
superscriptaddress,.
amsmath,
amssymb,
aps,
longbibliography,
floatfix,
]{revtex4-2}
\usepackage{mathrsfs,braket}
\usepackage{physics}
\usepackage{amssymb, amsbsy, amsmath, latexsym, dsfont, array, layout, color, mathtools}
\usepackage{amsfonts}
\usepackage{calc}
\usepackage[dvipsnames]{xcolor}
\usepackage{graphicx}
\usepackage{algorithm}
\usepackage{epstopdf}
\usepackage{bm}
\usepackage[normalem]{ulem}
\usepackage[unicode=true,breaklinks=true,colorlinks=true]{hyperref}
\usepackage{bigints}
\usepackage[T1]{fontenc}
\usepackage{newtxtext}
\usepackage[utf8]{inputenc}
\usepackage{upgreek}
\usepackage{siunitx}

\hypersetup{citecolor=blue,urlcolor=blue,linkcolor=blue, filecolor=blue}

\def\to{\rightarrow}

\def\ket#1{| #1 \rangle}

\def\to{\rightarrow}

\usepackage[normalem]{ulem}

\begin{document}

\title{Practical advantage of non-Hermitian enhanced quantum sensing}

\author{Kun Yang}
\author{Yaoming Chu}
\email{yaomingchu@hust.edu.cn}
\author{Musang Gong}
\email{musang_gong@hust.edu.cn}
\author{Ning Wang}
\email{ningwang@hust.edu.cn}
\author{Jianming Cai}
\affiliation{School of Physics, Center for Intelligence and Quantum Science (CIQS), International Joint Laboratory on Quantum Sensing and Quantum Metrology, Huazhong University of Science and Technology, Wuhan, 430074, China}
\affiliation{Hubei Key Laboratory of Gravitation and Quantum Physics, Institute for Quantum Science and Engineering, Huazhong University of Science and Technology, Wuhan, 430074, China}

\begin{abstract}
Non-Hermitian systems have emerged as a powerful paradigm for ultrasensitive sensing, leveraging unique spectral and dynamical properties unmatched in Hermitian physics. While recent theoretical bounds suggest these protocols offer no metrological advantage over Hermitian ones in the ideal shot-noise-limited regime—when rigorously accounting for the success probability of non-unitary evolution—their practical utility in realistic experimental conditions has not yet been systematically explored. In this work, we shift the focus toward practical laboratory performance and demonstrate that non-Hermitian sensing protocols can significantly outperform their Hermitian counterparts in the presence of pervasive classical technical noises. This performance gain mainly stems from a strongly enhanced susceptibility that amplifies the signal response, effectively overcoming the precision floor imposed by technical imperfections. By numerically evaluating the Fisher information under technical noise, we further substantiate the regimes where non-Hermitian platforms yield definitive practical gains. Our results reconcile the ongoing debate between fundamental limits and experimental observations, offering a concrete avenue for building high-precision, noise-resilient sensors.
\end{abstract}

\maketitle

\section{Introduction}
\label{sec:level1}

Over the past few decades, the study of non-Hermitian systems has burgeoned into a vibrant interdisciplinary frontier, profoundly reshaping our understanding of open quantum dynamics and wave phenomena~\cite{ashida2020non,bergholtz2021exceptional,el2018non}. A foundational cornerstone of this field was the pioneering discovery by Bender and Boettcher that non-Hermitian Hamiltonians—most notably those invariant under parity-time ($\mathcal{PT}$) symmetry—can possess entirely real spectra~\cite{bender1998real,Bender_2007,PhysRevLett.89.270401,PatrickDorey_2001}. This conceptual breakthrough shattered the long-held dogma that Hermiticity is an unyielding prerequisite for physical observables, thereby catalyzing a wealth of avenues spanning foundational quantum mechanics to diverse experimental explorations~\cite{heiss2000repulsion,doppler2016dynamically,kawabata2019symmetry}.  

At the heart of these developments is the manifestation of non-Hermitian spectral degeneracies known as exceptional points (EPs), where both eigenvalues and eigenvectors coalesce simultaneously~\cite{heiss2012physics,WDHeiss_2004}. In the vicinity of an $n$th-order EP, a weak perturbation of strength $\epsilon$ induces a sublinear spectral splitting scaling as $\epsilon^{1/n}$~\cite{jing2017high,wiersig2016sensors}. Because the corresponding susceptibility—the derivative of this spectral response—formally diverges in the limit of vanishing perturbations, EPs have ignited an explosion of interest in enhanced sensing protocols. This paradigm has been successfully manifested across a remarkably diverse spectrum of macroscopic coupled-mode models, including optical microresonators~\cite{feng2011nonreciprocal,ruter2010observation,chen2017exceptional,hodaei2017enhanced,feng2014single,doi:10.1126/science.aar7709,PhysRevApplied.12.024002,zhang2018phonon,PhysRevLett.117.123601,guo2009observation,ren2017ultrasensitive,Mortensen:18,PhysRevA.96.033842,PhysRevLett.108.024101,PhysRevLett.86.787,lai2019observation}, electric circuits~\cite{choi2018observation,zou2021observation,deng2024ultrasensitive} and mechanical systems~\cite{bender2013observation,li2024observation}. 

However, a profound caveat has emerged: this singular spectral response does not straightforwardly translate into enhanced metrological precision~\cite{wiersig2020prospects,bao2021fundamental,lau2018fundamental,zhang2019quantum,chu2020quantum,PhysRevLett.117.110802,langbein2018no,PhysRevLett.131.160801,yu2023quantum}. A fundamental bottleneck specific to EP-based sensing is rooted in the very mechanism that drives it—the coalescence of eigenvectors. As the eigenmodes become quasi-parallel, their distinguishability is severely degraded, a vulnerability that can completely nullify the benefits of an amplified susceptibility~\cite{wiersig2020prospects,chen2019sensitivity}. In optical mode systems like laser-based gyroscopes, this mode non-orthogonality triggers an excess noise penalty; it amplifies the coupling of stochastic fluctuations, such as spontaneous emission, into the lasing mode, severely broadening the spectral linewidth beyond the standard Schawlow-Townes limit. This degradation is formally quantified by the Petermann factor $K$~\cite{wang2020petermann, wiersig2016sensors,wiersig2020prospects,PhysRevA.101.053846}. Near an $n$th-order EP, the Petermann factor diverges as $K\propto \epsilon^{-2(n-1)/n}$, precisely canceling the sensitivity gain and restoring the conventional scaling of the overall signal-to-noise ratio. Beyond noise amplification, extending EP sensors to quantum-coherent platforms—such as solid-state spins~\cite{wu2019observation}, dissipative ultracold atoms~\cite{li2019observation}, and superconducting qubits~\cite{naghiloo2019quantum}—unveils a severe temporal constraint. Resolving the singular splitting $\Delta E \sim \epsilon^{1/n}$ imposes an interrogation time $T\sim 1/\Delta E$. This ``critical slowing down" forces the required measurement time to diverge as the perturbation vanishes. These fundamental physics limitations are further compounded by the stringent experimental demands of the non-Hermitian critical regime, which requires exquisite fine-tuning of system parameters such as the delicate balance of dissipation rates against coherent coupling and the precise calibration of driving field intensities. 

To circumvent these fundamental and operational hurdles—particularly the severe temporal constraints and the daunting requirement for experimental stabilization—the frontier of non-Hermitian metrology has recently expanded into broader, alternative paradigms beyond the traditional EP framework. One prominent avenue focuses on exploiting {\it non-Hermitian quantum dynamics} away from any spectral degeneracies, harnessing intrinsic features such as complex-valued eigenenergies and eigenstate non-orthogonality~\cite{chu2020quantum,PhysRevLett.133.180801,g3n3-gh49}. By deliberately operating in the non-degenerate regime, these dynamical protocols aim to boost parameter sensitivity while entirely bypassing the penalty of critical slowing down. In parallel, {\it non-Hermitian topology} has emerged as another powerful strategy. Here, phenomena such as the non-Hermitian skin effect leverage the extreme susceptibility of the bulk spectrum to boundary perturbations, providing a sensing mechanism that represents a radical departure from the localized Hermitian paradigms~\cite{PhysRevLett.128.223903,PhysRevLett.125.180403,zou2021observation}. 

Despite the conceptual richness of these strategies, the physical realization of non-Hermiticity remains a nontrivial task, given that closed quantum systems are fundamentally governed by Hermitian Hamiltonians.  A conventional route utilizes open quantum systems described by Lindblad master equations~\cite{PhysRevLett.68.580,PhysRevA.100.062131}. In the quantum-trajectory picture, conditioning the system’s evolution on the strict absence of quantum jumps yields an effective non-Hermitian Hamiltonian. However, this null-measurement approach suffers from a success probability that decays exponentially with time, and experimentally isolating these rare, jump-free trajectories presents a daunting task. A more controllable alternative engineers the target non-Hermitian dynamics within an enlarged, dilated Hermitian space via intentional postselection~\cite{PhysRevLett.101.230404,PhysRevLett.119.190401wu2019observation}. Yet, both methodologies are ultimately bound by the diminishing success probability of postselection. From a rigorous resource-theoretic perspective, once this probabilistic overhead is factored into the metrological accounting, non-Hermitian sensors typically yield no fundamental advantage over optimal Hermitian strategies, especially when evaluated under the metric of quantum Fisher information~\cite{Naikoo_2026,g3n3-gh49,qk5r-h851}.

Such a discouraging verdict, however, hinges on an idealized picture where precision is strictly governed by fundamental quantum fluctuations. In practical laboratory settings, however, the actual performance ceiling is often dictated by technical imperfections rather than these ultimate quantum bounds. This discrepancy leads to a pivotal question: does the lack of an intrinsic quantum advantage preclude the practical utility of non-Hermitian sensing in real-world applications? While previous studies have extensively scrutinized fundamental limits like shot noise, the impact of pervasive technical noise sources—such as readout noise, systematic calibration offsets, and detector saturation—has not yet been systematically explored. In a plethora of experimental scenarios, it is precisely these classical imperfections that constitute the dominant noise floor. Understanding how non-Hermiticity reconfigures the system's response to these practical constraints is therefore essential to determine whether such platforms can transition from theoretical curiosities to functional quantum technologies.

In this work, we bridge this gap by establishing a systematic theoretical framework to evaluate non-Hermitian sensing under diverse, experimentally relevant technical noise sources. We demonstrate that, while non-Hermitian protocols may lack a universal advantage in the ideal shot-noise-limited regime, they offer substantial noise resilience in realistic scenarios where classical imperfections cannot be efficiently mitigated by standard averaging over repetitive measurements. By invoking an information-theoretic approach based on Fisher information, we also identify the operational regimes where non-Hermitian sensors demonstrate clear metrological gains. Ultimately, our results provide a concrete roadmap for translating non-Hermitian physics into high-precision, noise-resilient quantum sensing technologies.

The remainder of this paper is structured as follows. In Sec.~\ref{sec:mechanisms}, we review the core mechanisms of non-Hermitian sensing, 
highlighting the distinctions between EP-based and alternative dynamical non-Hermitian approaches, alongside a comparison with the Hermitian counterparts. Sec.~\ref{sec:Advantage} elucidates the general principle under which non-Hermitian sensing protocols yield significant practical advantage. The impacts of various technical noise sources on the measurement precision are systematically scrutinized in Sec.~\ref{sec:noise}. 
We further establish an information-theoretic perspective based on Fisher information in Sec.~\ref{sec:information}. 
Finally, Sec.~\ref{sec:discussion} summarizes our findings and discusses the implications for experimental realization.

\section{Non-Hermitian Sensing}
\label{sec:mechanisms}

\subsection{Exceptional-point based sensing}
The use of non-Hermitian physics for quantum sensing can be traced back to the discovery that macroscopic coupled-mode systems operating near an exceptional point (EP) exhibit an anomalously amplified response to weak external perturbations \cite{wiersig2016sensors}. This behavior sharply contrasts with that of Hermitian systems, where the eigenvalue splitting scales linearly with the perturbation strength near ordinary degeneracies~\cite{heiss2012physics,berry2004physics}. The response enhancement near an EP arises from the coalescence of both eigenvalues and eigenvectors—a feature unique to non-Hermitian degeneracies—which gives rise to non-analytic spectral susceptibility. 

To grasp the essential mechanism underlying this enhancement, it is instructive to examine a minimal two-level non-Hermitian Hamiltonian,
\begin{equation}
H_0 =
\begin{pmatrix}
0 & A \\
A^\prime & 0
\end{pmatrix},
\end{equation}
where $A$ and $A^\prime$ are taken to be real parameters. This simple model successfully captures the qualitative distinction between Hermitian and non-Hermitian behavior. For $A=A^\prime$, the Hamiltonian is Hermitian, and the two eigenmodes remain orthogonal. Conversely, when $A\neq A^\prime$, the system breaks Hermiticity, causing the eigenmodes to lose their orthogonality and ultimately coalesce into a single defective state at the EP. In both scenarios, the unperturbed eigenenergies coincide at $E=0$; however, the underlying geometric structure of the degeneracy is fundamentally distinct.
To probe how these two scenarios respond to weak external perturbations, we incorporate a small symmetric coupling of strength $\lambda$,
\begin{equation}
H = H_0 + H', \qquad
H' = 
\begin{pmatrix}
0 & \lambda \\
\lambda & 0
\end{pmatrix},
\end{equation}
and examine the resulting spectral splitting. When $H_0$ is nearly degenerate, the behavior is strikingly different in the Hermitian and non-Hermitian cases:
\begin{align}
\Delta E_{\text{H}} &\simeq \lambda, \\
\Delta E_{\text{nH}} &\simeq \sqrt{\lambda}.
\end{align}
The Hermitian case exhibits a standard linear response given by first-order perturbation theory, whereas at the second-order EP the square-root dependence reveals a characteristic branch-point structure of non-Hermitian degeneracies. More broadly, an EP of order $n$ produces an eigenvalue splitting that scales as $\lambda^{1/n}$ for perturbation $\lambda$, implying an increasingly accentuated susceptibility to weak perturbations by orders of magnitude as the EP order increases. This striking scaling has inspired numerous theoretical proposals and experimental demonstrations aiming to exploit EPs for enhanced sensing beyond conventional paradigms~\cite{jing2017high,wiersig2016sensors}.

Here we focus on EP-based sensing in single quantum systems. Consider a single-qubit dissipative system described by the following Hamiltonian,
\begin{equation}
    \hat{H}_0= J\hat{\sigma}_x-i\frac{\Gamma}{2} \hat{\sigma}_z -i\frac{\Gamma}{2}\hat{\mathbb{I}},
    \label{barehami}
\end{equation}
where $\hat{\sigma}_{x,z}$ are the Pauli operators, $\hat{\mathbb{I}}$ represents the identity operator. If neglecting the last global decay term, Eq.\,\eqref{barehami} can be viewed as a PT-symmetric Hamiltonian and exhibits a second-order EP at $J=\Gamma/2$. Similarly, we assume that a perturbation term of $\hat{H}_s=\lambda \sigma_z$ that emulates a weak external signal is applied, which leads to a spectral splitting of $\Omega=[J^2+(\lambda-i \Gamma/2)^2]^{1/2}$. This splitting can be measured based on the following theoretical relations
\begin{equation}
    P_J(T)+P_\Gamma(T)=e^{-\Gamma T}\left| \sin (\Omega T) \right|^2,
\end{equation}
where
\begin{equation}
    \begin{aligned}
        P_J(T)&=\left| \langle \uparrow | U(T) | \downarrow \rangle \right|^2,\\
        P_\Gamma(T)&=\left|\langle - |U(T)|+\rangle\right|^2.
    \end{aligned}
\end{equation}
Here, $U(T)=\exp[-i (\hat{H}_0+\hat{H}_s) T]$ represents the non-Hermitian propagator and $|\pm\rangle = (|\uparrow\rangle \pm| \downarrow \rangle)/\sqrt{2}$ denote the eigenbasis along the $x$ direction.

However, one must be cautious in interpreting the enhanced spectral response near an EP as a guaranteed metrological advantage~\cite{wiersig2020prospects,chen2019sensitivity}. In the above protocol, the effective detection of $\Omega$ also requires a diverging time to ensure $\Omega T\sim O(1)$ as $\lambda \to 0$, which is indeed an impractical requirement in a real sensing process and suffers from severe decoherence effect. Next we present a more general framework which does not rely on EP, and non-Hermitian criticality can be accessed at finite evolution time.

\subsection{Dynamical non-Hermitian sensing framework}
Transcending conventional EP-based sensing schemes, non-Hermitian dynamics can also significantly enhance the susceptibility to a weak perturbation. To illustrate this mechanism, we consider a generic single-qubit non-Hermitian Hamiltonian denoted by $H_0$, the eigenvectors of which are generally non-orthogonal and can be represented by
\begin{equation}
\label{Eq:nonHEV}
\begin{aligned}
|\psi_{+}\rangle &\sim |0\rangle + \delta |1\rangle, \\
|\psi_{-}\rangle &\sim |0\rangle - (\delta/a) |1\rangle,
\end{aligned}
\end{equation}
where $a, \delta \in \mathbb{C}$ and $|a|\leq 1$ without loss of generality. Here, $\delta$ characterizes the deviation from the computational basis state $|0\rangle$, and $a$ controls the overlap between $|\psi_\pm\rangle$. By further assuming the two associated eigenvalues as $\mathcal{E}_+$ and $\mathcal{E}_-$ respectively, the Hamiltonian takes the following explicit form
\begin{equation}
H_0=\frac{\mathcal{E}_+-\mathcal{E}_-}{1+a}\begin{pmatrix} 1 & a\delta^{-1} \\ \delta & a \end{pmatrix}+\mathcal{E}_- I.
\label{Eq:nonHEV}
\end{equation}
For simplicity, we set \(\mathcal{E}_{+} = 2\mathcal{E} \in \mathbb{C}\) and \(\mathcal{E}_{-} = 0\).

Following the same procedure, we proceed to consider such a non-Hermitian qubit system as a sensor perturbed by an external weak signal $\lambda V$~\cite{chu2020quantum}, where $\lambda$ represents the unknown parameter to be estimated; the corresponding total Hamiltonian reads,
\begin{equation}
H = H_0 + \lambda V.
\end{equation}
For concreteness, here we take $V=\sigma_x$.  The eigenvectors of $H$ can still be reformulated as the general form of Eq.\,\eqref{Eq:nonHEV}, but with the parameters $\{\delta,a, \mathcal{E}\}$ implicitly depending on $\lambda$ hereafter. Starting from the initial state $\ket{\psi(0)}=\ket{0}$, the time-evolved state is given by
\begin{equation}
\label{Eq:nonHstate}
\begin{aligned}
\left | \psi(t) \right \rangle &=e^{-iHt}\left | \psi(0) \right \rangle\\
&=\frac{1-e^{i2\mathcal{E} t}}{1+a}\left( \mathcal{D}_t \left | 0 \right \rangle +\delta \left | 1 \right \rangle \right),
\end{aligned}
\end{equation}
where $\mathcal{D}_t$ determines the amplitude of the initial state during evolution
\begin{equation}
2\mathcal{D}_t=(1-a)+i(1+a)\cot({\mathcal{E} t}).
\end{equation}
Immediately, the probability of obtaining a measurement outcome $\left | 0 \right \rangle$ is given by
\begin{equation}
\label{Eq:ideal_PnH}
p_{\mathrm{nH}}(t,\lambda)=\frac{\left | \mathcal{D}_t \right |^2}{\left | \delta \right |^2+\left | \mathcal{D}_t \right |^2},
\end{equation}
Its susceptibility to  $\lambda$ is characterized by
\begin{equation}
\label{Eq:ideal_chi_nH}
\chi_{\mathrm{nH}} (\lambda) \equiv \frac{\partial p_{\mathrm{nH}}(t,\lambda)}{\partial \lambda} 
= \chi_{\left| \delta \right| }\, \frac{\partial \left|\delta\right|}{\partial \lambda}
+ \chi_{\left|\mathcal{D}_t\right|} \, \frac{\partial \left|\mathcal{D}_t\right|}{\partial \lambda},
\end{equation}
where $\chi_{|\delta|} = \partial p_{\mathrm{nH}}/\partial |\delta|$, and $\chi_{|\mathcal{D}_t|} = \partial p_{\mathrm{nH}}/\partial |\mathcal{D}_t|$. When $|\mathcal{D}_t|\sim|\delta|\ll 1$, a small change of $\lambda$ can lead to a change of $|\mathcal{D}_t|$ and $|\delta|$ on the same order as their own magnitudes, thereby giving rise to a significant change of $p_{\mathrm{nH}}(t,\lambda)$. This feature indicates an extreme sensitivity of  $p_{\mathrm{nH}}(t,\lambda)$ to $\lambda$.

\begin{figure}
    \centering
    \includegraphics[width=0.98\linewidth]{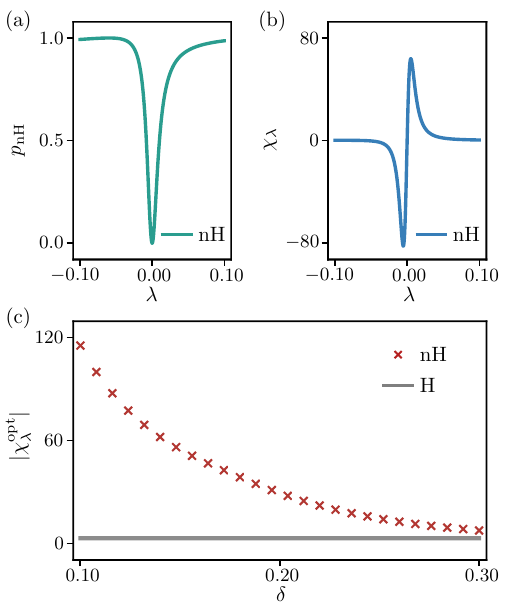}
    \caption{{\bf Non-Hermitian sensing mechanism.} (a) The probability $p_{\mathrm{nH}}(t,\lambda)$ [Eq.\,\eqref{Eq:ideal_PnH}] as a function of the parameter $\lambda$ to be estimated. (b) The associated susceptibility $\chi_{\mathrm{nH}} (\lambda)$ [Eq.\,\eqref{Eq:ideal_chi_nH}]. (c) Optimized susceptibility in (b) maximized over $\lambda$. In the regime $|\mathcal{D}_t|\sim|\delta|\ll 1$,  $p_{\mathrm{nH}}(t,\lambda)$ exhibits a sharply divergent response to $\lambda$ as $|\delta|\to 0$, markedly outperforming the conventional Ramsey interferometry in Eq.\,\eqref{Eq:ideal_PH}. The gray line indicates the optimal susceptibility $\chi_{\mathrm{H}}^{\mathrm{max}} = t$ of the latter, serving as the Hermitian baseline. Parameters for the sensing protocol are set to $a=1$, $\mathcal{E}=0.5$, $t=\pi/(2\mathcal{E})$, with $\delta=0.12$ used in (a).}
    \label{fig_chi}
\end{figure}

As a comparison, we also present the result of the standard Hermitian scenario where $H=\lambda \sigma_x$. Accordingly, the probability of the time-evolved system in state $\left | 0 \right \rangle$ can be expressed as
\begin{equation}
\label{Eq:ideal_PH}
    p_{\mathrm{H}}(t,\lambda)=\frac{1}{2}\left( 1+\mathrm{cos}2\lambda t \right),
\end{equation}
from which one immediately obtains the associated susceptibility as $\chi_{\mathrm{H}}(\lambda)  = \partial p_{\mathrm{H}}(t,\lambda)/ \partial \lambda=-t\sin(2\lambda t)$. From Fig.\,\ref{fig_chi}, it can be seen that for small values of $\delta$,
\begin{equation}
\label{Eq:chiRatio}
    \mathcal{A} = \left|\frac{\chi_{\mathrm{nH}}(\lambda)}{\chi_{\mathrm{H}}(\lambda)}\right|\gg 1.
\end{equation}
However, this does not simply imply that the non-Hermitian sensing scheme would outperform the Hermitian counterpart. 

\section{Advantage of Non-Hermitian Sensing}
\label{sec:Advantage}
In an explicit sensing protocol, the measurement precision of $\lambda$ using the formula of error-propagating can be expressed as
\begin{equation}
\delta \lambda = \frac{\sigma}{\chi_\lambda},
\end{equation}
where $\sigma$ characterizes the measurement error of the probability in state $|0\rangle$, and $\chi_\lambda$ is chosen as $\chi_{\mathrm{nH}}(\lambda)$ or $\chi_{\mathrm{H}}(\lambda)$ depending on the sensing scheme that is exploited. In actual experiments, different categories of noise can contribute to $\sigma$. For clarity, we decompose $\sigma$ into two components by considering its scaling relation with the effective measurement repetitions, denoted by $N$,
\begin{equation}
\label{Eq:noiseDecomposition}
\sigma^2=\sigma_{\mathrm{SQL}}^2+\sigma_{\mathrm{nSQL}}^2,
\end{equation}
where $\sigma_{\mathrm{SQL}} = \Theta(1/\sqrt{N})$ captures statistical scaling of errors that follows the central limit theorem, referred to as the standard quantum limit (SQL) or shot noise in quantum optics. The second term $\sigma_{\mathrm{nSQL}}$ represents technical noise that cannot be {\it efficiently} suppressed by averaging through repetitive experiments. Here, the symbol $\Theta(x)$ indicates an asymptotic scaling identical to $x$ for large $x$. 

For a non-Hermitian sensing protocol, the effective dynamics are typically embedded into a larger, dilated Hilbert space to be unitarily realized, e.g. via the Naimark dilation. The generation of the conditional state is inherently probabilistic, often bounded by a small postselection success probability. Following the dilation framework established for non-Hermitian systems~\cite{wu2019observation, PhysRevLett.123.080404}, here we consider a global Hermitian Hamiltonian $H_{\mathrm{tot}}$ acting on an extended Hilbert space, $\mathcal{H}_{\mathrm{sys}}\otimes \mathcal{H}_{\mathrm{anc}}$. A standard construction parametrizes the total Hamiltonian as
\begin{equation}
\label{Eq:DilationHtot}
    H_{\text{tot}} = \Lambda\otimes \mathbb{I} + \Gamma\otimes\sigma_z + \lambda\sigma_z\otimes \mathbb{I},
\end{equation}
where $\sigma_z$ denotes the standard Pauli-Z operator, while $\Lambda$ and $\Gamma$ represent Hermitian operators defined on the system space $\mathcal{H}_{\mathrm{sys}}$. These operators are constructed from the non-Hermitian Hamiltonian $H_0$ in Eq.\,\eqref{Eq:nonHEV} and a metric-associated operator $M=\mathrm{diag}[1,a/\delta^2]$ via the relations 
\begin{equation}
\begin{aligned}
    \Gamma = i(H_{0}U - UH_{0})M^{-1}, \\
    \Lambda = (H_{0}+UH_{0}U)M^{-1}.
\end{aligned}
\end{equation}
with $U=\sqrt{M-\mathbb{I}}$. Crucially, this dilated formulation ensures that the interaction with the unknown parameter
$\lambda$ retains a linear, standard coupling form, enabling a direct metrological comparison with conventional Hermitian sensing protocols. Starting from an initial product state $\ket{\Psi(0)}=|0\rangle \otimes (|0\rangle-i|1\rangle)/\sqrt{2}$, the global unitary evolution $\ket{\Psi(t)}=e^{-i H_{\mathrm{tot}} t}\ket{\Psi(0)}$ can be partitioned in the system-ancilla basis as
\begin{equation}
\label{Eq:DilatedState}
\ket{\Psi(\lambda, t)} = \ket{\psi (\lambda, t)}\otimes\left | - \right \rangle+\ket{\zeta(\lambda, t)}\otimes\left | + \right \rangle,
\end{equation}
where $|\pm\rangle$ denote the eigenstates of the ancilla subspace along the $x$ direction. Here, the unnormalized component $|\psi(\lambda, t)\rangle$ corresponds to the state evolved under an effective non-Hermitian Hamiltonian. The corresponding success probability of projecting the ancilla into the target state $|-\rangle$ is given by the squared norm of this conditional state,
\begin{equation}
P_{\mathrm{suc}}(\lambda) = \|( I\otimes\left | - \right \rangle\left \langle - \right | )\ket{\Psi(\lambda, t)}\|^2.
\end{equation}
This probability of implementing a desired non-Hermitian dynamics via the dilation protocols into an enlarged Hermitian system is usually very small for $|\delta|\ll 1$. In fact, the product $P_{\mathrm{suc}}(\lambda) \chi_{\mathrm{nH}}^2(\lambda)\sim P_{\mathrm{suc}}(\lambda) \mathcal{A}^2$ is often approximately equal to a constant \cite{chu2020quantum}. Consequently, the effective number of available measurement trials for the non-Hermitian case would be much smaller than the Hermitian one under an identical number of total repetitive measurements $N_{\mathrm{t}}$, i.e. $N_{\mathrm{nH}}\ll N_{\mathrm{H}} = N_{\mathrm{t}}$.

First, we consider the simple case where $\sigma_{\mathrm{nSQL}}=0$. Using the scaling relation that $N_{\mathrm{nH}} =\Theta(N_{\mathrm{t}} / \mathcal{A}^2)$, the precision achieved by the non-Hermitian scheme is indeed comparable to the Hermitian counterpart,
\begin{equation}
\label{Eq:SQLcase_precision}
 \delta \lambda_{\mathrm{nH}} = \frac{\sigma_{\mathrm{SQL}}}{\chi_{\mathrm{nH}}(\lambda)} \sim \frac{1/\sqrt{N_{\mathrm{t}}/\mathcal{A}^2}}{\mathcal{A}\chi_{\mathrm{H}}(\lambda)} \sim  \frac{1/\sqrt{N_{\mathrm{t}}}}{\chi_{\mathrm{H}}(\lambda)} \sim \delta \lambda_{\mathrm{H}}.
\end{equation}
This result is consistent with the conjecture through the perspective of quantum Fisher information of the full quantum system~\cite{chu2020quantum,PhysRevLett.131.160801}. 

Surprisingly, albeit the small success probability, the particular subensemble of the total measurement outcomes that yields the desired non-Hermitian dynamics is sufficient to achieve a precision of the same order as the full system. This feature indeed endows the non-Hermitian strategy a peculiar capability of suppressing classical technical noise. Next, we explicitly focus on the realistic case that $\sigma_{\mathrm{nSQL}}\neq 0$. Suppose that $\sigma_{\mathrm{nSQL}}=\Theta (N^{-\alpha})$ with $\alpha <1/2$, which is much larger than $\sigma_{\mathrm{SQL}}=\Theta (N^{-1/2})$ in the large-$N$ limit. Similarly, we find that
\begin{equation}
\label{Eq:nHadvantage}
 \delta \lambda_{\mathrm{nH}} \sim \frac{(N_{\mathrm{t}}/\mathcal{A}^2)^{-\alpha}}{\chi_{\mathrm{nH}}(\lambda)} = \Theta \left( \mathcal{A}^{2\alpha-1} \right) \delta \lambda_{\mathrm{H}}.
\end{equation}
This formula clearly uncovers the significant regime of non-Hermitian sensing to reduce experimental technical noise that can not be efficiently eliminated by repetitive measurements.

\section{Technical noise models}
\label{sec:noise}
In this section, we explicitly analyze different technical noise sources that contribute to the error term $\sigma_{\mathrm{nSQL}}$ in Eq.\,\eqref{Eq:noiseDecomposition} that does not follow the central limit theorem. Specifically, we focus on common experimental imperfections, including biased readout, background detection noise, systematic parameter offsets, and detector saturation. By studying these explicit noise models, we establish the practical advantage under which non-Hermitian sensing architectures definitely achieve a better precision than conventional Hermitian protocols.

\subsection{Biased State Readout}
In state-of-the-art quantum sensing platforms, ranging from superconducting circuits to nitrogen-vacancy (NV) centers, readout fidelity imposes a primary bottleneck for the achievable measurement precision. Beyond intrinsic 
quantum projection noise, the physical process of state discrimination often introduces an irreducible systematic shift that persists regardless of the number of measurement repetitions $N$. This constant bias originates from inherent physical asymmetries during the readout window, driving the observed signal away from its true value.

\begin{figure}[t]
\centering
\includegraphics[width=0.95\linewidth]{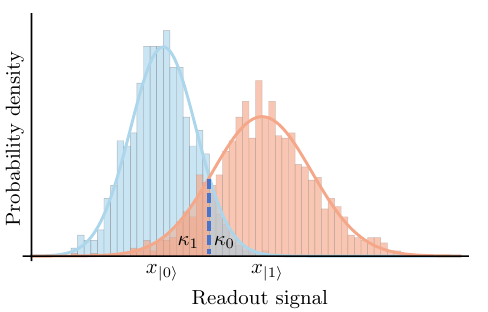}
\caption{{\bf Readout of a generic qubit sensor.} Histograms of detected photon counts for a qubit prepared in the $\ket{0}$ (blue) and $\ket{1}$ (orange) states. Owing to the photon shot noise and classical readout imperfections, the empirical outcomes follow broadened distributions centered at $x_{\left | 0 \right \rangle}$ and $x_{\left | 1 \right \rangle}$. The finite overlap between the profiles leads to misassignment errors, characterized by the probabilities $\kappa_0$ and $\kappa_1$. Crucially, imprecise calibration of these error rates yields an unknown systematic measurement bias in the population, which cannot be circumvented via repetitive statistical averaging.}
\label{fig:readout_error}
\end{figure}

To quantify this, we consider a generic qubit sensor subjected to $N$ repeated measurements following the parameter interrogation stage, yielding counts $N_0$ and $N_1$ for the states $\ket{0}$ and $\ket{1}$, respectively (see Fig.\,\ref{fig:readout_error}). In an ideal scenario, the measurement probability,
\begin{equation}
p_0 = \frac{N_0}{N_0+N_1},    
\end{equation}
converges to the theoretically predicted projection in the large-$N$ limit. In practice, however, hardware-specific imperfections—specifically the competition between readout dynamics and the internal relaxation of the qubit—distort this distribution. In superconducting transmons or trapped ions, for instance, longitudinal relaxation ($T_1$) during the dispersive readout pulse causes the excited state $\ket{1}$ (often the ``bright" state) to decay into the ground state $\ket{0}$ (the ``dark" state). Conversely, the upward transition $\ket{0} \to \ket{1}$ is thermally suppressed at millikelvin temperatures~\cite{PhysRevApplied.7.054020,PhysRevLett.100.200502,chen2023transmon}. This mechanism results in an asymmetric error hierarchy where the ``dark" error rate ($\kappa_1$, misidentifying $\ket{1}$ as $\ket{0}$) significantly outweighs the ``bright" error rate ($\kappa_0$, misidentifying $\ket{0}$ as $\ket{1}$), systematically biasing the population toward the ground state. For NV centers in diamond, the manifestation of this bias is reversed but follows an analogous physical logic. The ground-state manifold is a spin triplet, where the $m_s=0$ state (denoted as $\ket{0}$) and the $m_s=\pm 1$ state ($\ket{\pm1}$) exhibit distinct optical properties under continuous green laser excitation. The bright state $\ket{0}$ undergoes spin-preserving cycling transitions to emit high photoluminescence, whereas the dark states $\ket{\pm1}$ preferentially decay into metastable singlets via non-radiative inter-system crossing~\cite{DOHERTY20131,Robledo_2011}. Crucially, this measurement paradigm induces an intrinsic spin polarization within a typical $300$-$500$ ns readout window: the shelved population asymmetrically repopulates the $\ket{0}$ state upon leaving the singlet manifold. This unidirectional $\ket{\pm1} \to \ket{0}$ state conversion, compounded by overlapping Poissonian photon-counting statistics and fixed hardware thresholding, leads to a readout error hierarchy.

In both physical implementations, these state-dependent readout errors map the ideal projection probability into an empirically observed probability,
\begin{equation}
p_0^\prime = p_0 (1-\kappa_0) + (1-p_0)\kappa_1,
\end{equation}
where $\kappa_0$ and $\kappa_1$ are misidentification rates of state $\ket{0}$ or $\ket{1}$, respectively. This mapping introduces a systematic, state-dependent measurement bias,
\begin{equation}
    \text{Bias}(p_0)=\kappa_1-(\kappa_0+\kappa_1)p_0,
\end{equation}
which represents an irreducible shift that cannot be completely mitigated by standard statistical averaging when a priori exact knowledge of $\kappa_{0,1}$ is unavailable.  In this scenario, the utility of non-Hermitian sensing becomes manifestly evident: by markedly amplifying the response function $\chi_{\mathrm{nH}}(\lambda)$, the sensor can enhance the signal magnitude far above this hardware-inherent bias. This amplification effectively shields the measurement from the deleterious effects of asymmetric readout noise, restoring high-fidelity sensing in the presence of intrinsically biased distributions.


\subsection{Correlated Background Noise}
Beyond state readout bias, quantum sensors are frequently subjected to background detection noise that exhibits significant temporal or spatial correlations. In contrast to uncorrelated stochastic fluctuations, which are suppressed in the large-$N$ limit through averaging, correlated noise introduces a fundamental precision floor that cannot be eliminated by increasing the number of repetitive measurements. We consider a general measurement model in which the detected signal is given by
\begin{equation}
 X_j = x_j + \xi_j,   
\end{equation}
where $x_j$ denotes the intrinsic qubit response and $\xi_j$ represents the background detection noise. We assume that $x_j$ follows a mixture of two Gaussian distributions,
\begin{equation}
\label{Eq:phtonGaussian}
x_j\sim p_0\mathcal{N}(x_{\left | 0 \right \rangle},\sigma^2)+(1-p_0)\mathcal{N}(x_{\left | 1 \right \rangle},\sigma^2),
\end{equation}
whose mean and variance are 
\begin{equation}
 \begin{aligned}
\mathbb{E}[x_j] &= p_0 x_{\left | 0 \right \rangle} + (1-p_0) x_{\left | 1 \right \rangle},\\
\operatorname{Var} [x_j] & = p_0(1-p_0) x^2+ \sigma^2,
\end{aligned}
\end{equation}
where $x = x_{\ket{0}} - x_{\ket{1}}$ defines the signal contrast.

In this correlated regime, the background detection noise $\{\xi_j\}_{j=1}^N$ is modeled as a multivariate Gaussian distribution with zero mean and covariance matrix $\mathbf{C}$, namely
\begin{equation}
\label{Eq: cbn_xi}
P(\xi) = \frac{1}{\sqrt{(2\pi)^N \det \mathbf{C}}}
\exp\!\left[-\frac{\xi^T \mathbf{C}^{-1}\xi}{2}\right].
\end{equation}
To quantify the influence of these correlations, we consider the following probability estimator 
\begin{equation}
\label{Eq: bdn_estimator}
\hat{p} = \frac{1}{Nx}\sum_j  (X_j - x_{\ket{1}}).
\end{equation}
While temporarily neglecting biased readout errors for clarity, one readily obtains $\mathbb{E}[\hat{p}]=p_0$, demonstrating $\hat{p}$ as an unbiased estimator of the population in state $|0\rangle$.

The performance of the sensor is determined by the variance of this estimator, 
\begin{equation}
\begin{aligned}
    \operatorname{Var}[\hat{p}]
    &=\frac{1}{N^2x^2}\sum_{ij}\mathbb{E}[(X_i-\mathbb{E}[X_i])(X_j-\mathbb{E}[X_j])] \\ 
    &=\frac{1}{N^2x^2}\sum_{i}\operatorname{Var}[x_i] + \frac{1}{N^2x^2}\sum_{ij}\mathbb{E}[\xi_i\xi_j]\\
    &=\frac{p_0(1-p_0)}{N}+\frac{\sigma^2}{Nx^2}+\frac{1}{N^2x^2}\sum_{ij} \mathbb{E}[ \xi_i\xi_j].
\end{aligned}
\end{equation}
In accordance with the noise decomposition in Eq.\,\eqref{Eq:noiseDecomposition}, the terms scaling as $1/N$ constitute the standard SQL contribution, 
\begin{equation}
\sigma^2_{\mathrm{SQL}} = \frac{p_0(1-p_0)}{N}+\frac{\sigma^2}{Nx^2},
\end{equation}
originating from the quantum projection noise and photon shot noise, respectively. Crucially, the influence of correlated background detection noise is quantified by the last term,
\begin{equation}
 \sigma^2_{\mathrm{nSQL}} = \frac{1}{N^2x^2}\sum_{ij} \mathbb{E}[ \xi_i\xi_j] = \frac{1}{N^2x^2}\sum_{ij} \mathbf{C}_{ij}.
\end{equation}
In many realistic detection processes, the background noise exhibits translational invariance in either the temporal or spatial domain, implying that the covariance matrix depends solely on the separation between measurements, i.e. $\mathbf{C}_{ij}=\mathcal{C}(|i-j|)$. 

Generally, $\mathcal{C}(|i-j|)$ decays monotonically as the interval $|i-j|$ increases. If this decay is sufficiently rapid—such as an exponential decay or an algebraic power law with a large exponent—the sum $\sum_{j} \mathcal{C}(|i-j|)$ converges to a constant value independent of $N$. In this short-range correlated limit, the total sum $\sum_{ij} \mathbf{C}_{ij}$ scales linearly with $N$. The noise thus behaves effectively as white noise, retaining the standard $1/N$ suppression in the estimator variance. Conversely, for systems plagued by long-range correlations or $1/f$-type noise, the sum scales super-extensively as $\sum_{ij} \mathbf{C}_{ij}\sim N^\beta$ with $\beta\in (1,2]$. This scaling behavior dictates that $\alpha = 1-\beta/2$ in Eq.\,\eqref{Eq:nHadvantage}, signaling a clear metrological advantage offered by non-Hermitian sensing protocols. In the extreme limit of perfect correlation, where $\mathbf{C}_{ij}=\sigma_\xi^2$ for all $i$, $j$ (corresponding to $\beta\to 2$), the background noise establishes a completely rigid, irreducible noise floor. This floor cannot be mitigated by increasing the number of measurement repetitions $N$, thereby imposing a fundamental precision ceiling on conventional Hermitian protocols. On the other hand, non-Hermitian sensing mechanisms can overcome this bottleneck by amplifying the response function $\chi_{\mathrm{nH}}(\lambda)$, effectively elevating the parameter-induced signal well above the persistent noise floor.

To be more explicit, we proceed to investigate the scaling behavior of $\sum_{ij} \mathbf{C}_{ij}$ as a function of $N$, defining the total correlation sum as
\begin{equation}
    S_N = \sum_{i=1}^{N} \sum_{j=1}^{N} \mathbf{C}_{ij}.
\end{equation}
Since the number of index pairs with a fixed separation \( d = |i - j| \) is exactly \( 2(N - d) \) for \( d \ge 1 \), the sum can be rewritten as
\begin{equation}
\begin{aligned}
    S_N &= N\mathcal{C}(0) + 2 \sum_{d=1}^{N-1} (N - d) \mathcal{C}(d) \\
    &= N\mathcal{C}(0) + 2N \sum_{d=1}^{N-1} \mathcal{C}(d) - 2 \sum_{d=1}^{N-1} d \, \mathcal{C}(d).
\end{aligned}
\end{equation}
Suppose that the correlation function decays algebraically as $\mathcal{C}(d) \sim c \, d^{-\gamma}$ with $\gamma>0$. For the long-range regime \( 0 < \gamma < 1 \), the asymptotic behavior of \( S_N \)  in the large-$N$ limit is governed by the last two terms \( N \sum_{d=1}^{N-1} d^{-\gamma} \) and \( \sum_{d=1}^{N-1} d^{1-\gamma} \), both scaling as \( N^{2-\gamma} \). This immediately yields
\[
S_N = \Theta( N^{2-\gamma}).
\]
For the marginal case \( \gamma = 1 \), the summation introduces a logarithmic correction
\[
S_N = \Theta( N \log N).
\]
By contrast, in the short-range regime \( \gamma > 1 \), $S_N$ is dominated by the first term and scales linearly with the system size,
\[
S_N \sim N.
\]
Consequently, for algebraically decaying correlations, the scaling exponent \( \beta \) defined via \( S_N \sim N^{\beta} \) is given by
\[
\beta =
\begin{cases}
2 - \gamma, & 0 < \gamma < 1, \\
1, & \gamma > 1.
\end{cases}
\]
This result explicitly demonstrates that short-range correlated background noise (\( \gamma > 1 \)) remains extensive, preserving the standard \( 1/N \) scaling of the estimator variance. Conversely, long-range correlations (\( 0 < \gamma < 1 \)) produce super-extensive noise accumulation, \( S_N \sim N^\beta \) with \( 1 < \beta < 2 \). Specifically, substituting this into the relation $\alpha = 1 - \beta/2$, we obtain
\[
\alpha =
\begin{cases}
\gamma/2, & 0 < \gamma < 1, \\
1/2, & \gamma > 1.
\end{cases}
\]
This scaling relation, in conjunction with the framework established in Eq.\,\eqref{Eq:nHadvantage}, clearly illustrates that non-Hermitian sensing offers a substantial performance advantage over conventional Hermitian protocols when operating in regimes plagued by long-range correlated background detection noise.

\begin{figure}[t]
\centering
\includegraphics[width=0.95\linewidth]{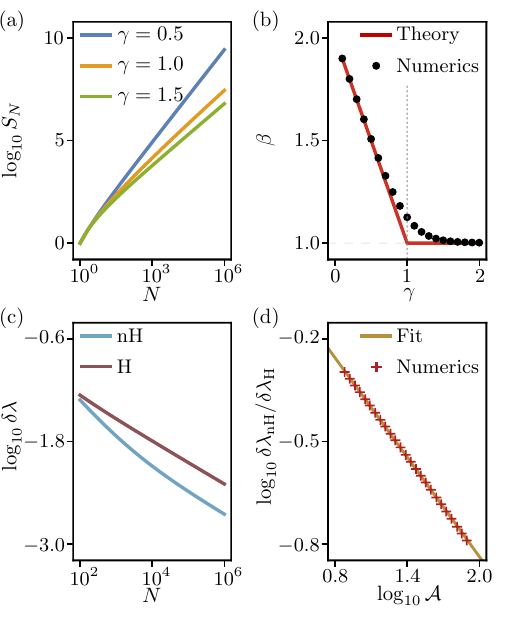}
\caption{{\bf Non-Hermitian sensing advantage under power-law correlated detection noise.}
(a) Total accumulated correlation
$S_N=\sum_{i,j=1}^{N}C_{ij}$
as a function of the measurement number $N$ for distinct correlation exponents $\gamma$. Here, the covariance matrix scales algebraically as $\mathcal{C}(|i-j|)\propto |i-j|^{-\gamma}$. In the short-range regime ($\gamma>1$), the accumulated correlation scales extensively,
$S_N\propto N$, whereas it grows super-extensively as $S_N\propto N^\beta$ with $\beta=2-\gamma$ in the long-range regime ($0<\gamma<1$).
(b) Extracted scaling exponent $\beta$ as a function of $\gamma$. The numerical results are well captured by the theoretical predictions: 
$\beta=2-\gamma$ for $0<\gamma<1$ and $\beta = 1$ for $\gamma>1$. 
(c) Measurement precision $\delta\lambda$ as functions of $N$ under a representative long-range correlated noise regime ($\gamma=0.5$). Largely enhanced parameter susceptibility ($|\chi_\mathrm{nH}|\simeq32.05$ at the working point $\lambda=-0.1279$) allows the non-Hermitian protocol to consistently maintain a lower uncertainty than the Hermitian baseline in the large-$N$ limit.
(d) Precision ratio between the non-Hermitian and Hermitian protocols as a function of $\mathcal{A}$ [Eq.\,\eqref{Eq:chiRatio}]. The variation of $\mathcal{A}$ is achieved by varying $\delta$ with $\chi_{\mathrm{nH}}(\lambda)$ maximized over $\lambda$. The fitted slope (brown line) of $-0.484$ is in close agreement with the theoretical prediction of $2\alpha-1=\gamma-1=-0.5$ according to Eq.\,\eqref{Eq:nHadvantage}. Parameters of non-Hermitian sensing based on the dilation framework in Eq.\,\eqref{Eq:DilationHtot} are set to $a=1$, $\mathcal{E}=0.5$, $t=\pi/(2\mathcal{E})$ and  $\delta=0.225$ in (c) and (d).}
\label{ling0}
\end{figure}

\subsection{Detector Saturation}

\begin{figure}
    \centering
    \includegraphics[width=0.95\linewidth]{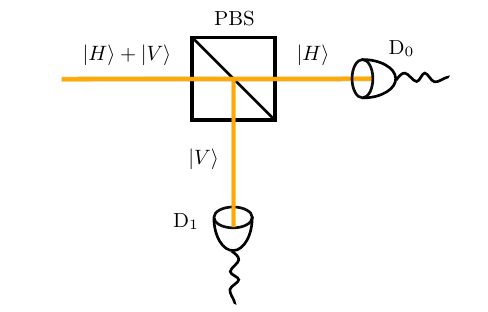}
    \caption{{\bf The paradigm of polarization-based photon detection}. A polarizing beam splitter (PBS) maps the orthogonal photon polarization states ($\left| H\right\rangle, \left| V\right\rangle$) onto distinct spatial modes monitored by photodetectors $D_0$ and $D_1$.}
    \label{picture2}
\end{figure}

The saturation effect of the photon detector is also a common limitation frequently encountered in optical measurements. In this subsection, we adopt a simplified two-detector description (see Fig.\,\ref{picture2}). Let $n_0$ and $n_1$ denote the photon numbers incident on detector $D_0$ and detector $D_1$, respectively, and $k_0$ and $k_1$ be the respective electrical outputs, the experimentally relevant estimator for the population associated with detector $D_0$ is then defined as
\begin{equation}
\label{Eq: electrProb}
    S=\frac{k_0}{k_0+k_1}.
\end{equation}
Within this framework, the photon counts registered by the two detectors are modeled as independent Poisson random variables,
\begin{equation}
\label{Eq: ds_photons}
\begin{aligned}
n_0 &\sim \mathrm{Poisson}\!\left[p(\lambda)n_{\mathrm{eff}}\right],\\
n_1 &\sim \mathrm{Poisson}\!\left[(1-p(\lambda))n_{\mathrm{eff}}\right],
\end{aligned}
\end{equation}
where $p(\lambda)$ denotes theoretical projection probability of the sensor onto the state $|0\rangle$ and $n_{\mathrm{eff}}$ denotes the total effective photon number preserved after the postselection stage. Specifically, for the non-Hermitian sensing protocol, this resource scales as $n_{\mathrm{eff}} = P_{\mathrm{suc}}(\lambda) n_{\mathrm{in}}$, where $n_{\mathrm{in}}$ is the total input photon budget and $P_{\mathrm{suc}}(\lambda)$ the parameter-dependent postselection success probability. Conversely, for the conventional Hermitian baseline, no postselection is performed, and the effective photon number reduces to the full initial input resource ($n_{\mathrm{eff}} = n_{\mathrm{in}}$).

For each detection channel $i\in\{0,1\}$, the distribution of output signal $k_i$  conditioned on the stochastically incident photon number $n_i$ is modeled as a discrete Gaussian distribution,
\begin{equation}
    P(k_i|n_i)=
    \frac{
    e^{-(k_i-\bar{k}_i)^2/2\sigma_{\mathrm{r}}^2}
    }{
    \sum_{k'_i} e^{-(k'_i-\bar{k}_i)^2/2\sigma_{\mathrm{r}}^2}
    }.
\end{equation}
Here, $\sigma_{\mathrm{r}}$ characterizes the electronic readout noise, and the conditional mean of the output signal is governed by the nonlinear transfer function \cite{PhysRevLett.118.070802}
\begin{equation}
    \bar{k}_i\equiv\bar{k}_i(n_i)
    =
    k_{\mathrm{m}}\!\left(1-e^{-n_i/n_{\mathrm{sat}}}\right),
\end{equation}
where $k_{\mathrm{m}}$ denotes the maximum attainable detector output and $n_{\mathrm{sat}}$ represents the characteristic saturation scale. This formulation smoothly captures the phenomenological behavior of detector saturation: the response scales sublinearly with the incident photon flux and approaches the asymptotic ceiling $k_{\mathrm{m}}$ in the large-$n_i$ limit.

Numerically, the mean $\mathbb{E}[S]$ and variance $\mathrm{Var}[S]$ of the probability estimator defined in Eq.\,\eqref{Eq: electrProb} are evaluated by generating a large ensemble of Monte Carlo trials according to this compound stochastic process. Consequently, the parameter estimation precision $\lambda$ is given by the standard error propagation formula,
\begin{equation}
    \label{Eq:ds_precision}
    \delta \lambda = \frac{\sqrt{\mathrm{Var}[S]}}{|\partial_\lambda \mathbb{E}[S]|}.
\end{equation}
By invoking the total derivative, the parametric susceptibility in the denominator can be explicitly expanded as
\begin{equation}
\partial_\lambda \mathbb{E}[S] = \frac{\partial \mathbb{E}[S]}{\partial p(\lambda)}\frac{\partial p(\lambda)}{\partial \lambda} + \frac{\partial \mathbb{E}[S]}{\partial P_{\mathrm{suc}}(\lambda)}\frac{\partial P_{\mathrm{suc}}(\lambda)}{\partial \lambda}.
\end{equation}
For the conventional Hermitian protocol, the second term vanishes identically since no postselection occurs ($P_{\mathrm{suc}}=1$). For the non-Hermitian protocol, operating deep within the postselected regime i.e. $P_{\mathrm{suc}}(\lambda)\ll 1$, this second term is also negligible since $\partial P_{\mathrm{suc}}(\lambda)/\partial \lambda $ is much smaller than the intrinsic parametric sensitivity of the theoretical projection probability $\chi (\lambda)=\partial p(\lambda)/\partial \lambda$.

\begin{figure}[t]
    \centering
    \includegraphics[width=0.95\linewidth]{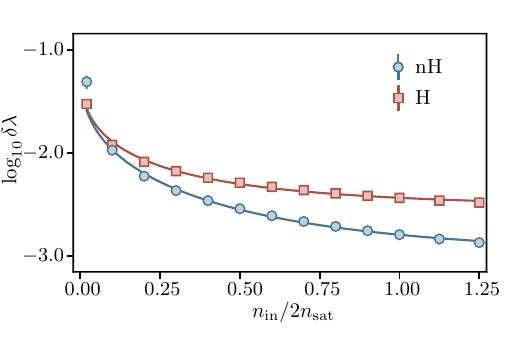}
    \caption{{\bf Non-Hermitian sensing advantage under detector saturation.}  As can be seen, non-Hermitian mechanism mitigates the performance degradation induced by detector clipping, yielding a lower precision floor compared to the Hermitian baseline. This advantage is achieved by compressing parameter information into a significantly smaller postselected photon ensemble, thereby preventing detector saturation. Here, photon numbers are sampled from Poisson distributions and converted to electronic outputs incorporating electronic readout noise. The parameter estimation precision is determined via error propagation [Eq.\,\eqref{Eq:ds_precision}] with the slope of the population estimator $S$ in Eq.\,\eqref{Eq: electrProb} estimated under a finite-difference step of $\Delta\lambda=10^{-3}$. Solid curves represent empirical fits modeled as $\delta\lambda(x)=A+B\exp[-(x/x_0)^{\beta}]$, where $x=n_{\mathrm{in}}/2n_{\mathrm{sat}}$ and $n_{\mathrm{in}}$ is the total input photon budget. For the numerical simulations, each data point is averaged over ten independent runs, with each run containing $2\times10^5$ random samples. Here, the non-Hermitian sensing protocol operates with $a=1$, $\mathcal{E}=0.5$, $t=\pi/(2\mathcal{E})$, $\delta=0.12$ at the working point $\lambda=-0.005$, under electronic noise parameter $n_{\mathrm{sat}}=1000$, $k_{\mathrm{m}}=4000$ and $\sigma_{\mathrm{r}}=10$. The fitted parameters are \((A,B,x_0,\beta)=(9.80\times10^{-4},\,8.14\times10^{-2},\,1.23\times10^{-2},\,0.360)\) for the non-Hermitian protocol and \((2.88\times10^{-3},\,7.90\times10^{-2},\,1.23\times10^{-2},\,0.348)\) for the Hermitian baseline.}
    \label{fig_SAT}
\end{figure}

Following an analysis analogous to Eq.\,\eqref{Eq:SQLcase_precision}, the intrinsic statistical scaling factors $\sqrt{\mathrm{Var}[S]}/\chi (\lambda)$ remain of the same order of magnitude for both sensing protocols. However, a stark distinction emerges via the first partial derivative, $\partial \mathbb{E}[S]/\partial p(\lambda)$, when the initial input resource $n_{\mathrm{in}}$ is comparable to the saturation scale $n_{\mathrm{sat}}$. The severe sublinear clipping of the detectors heavily degrades such a derivative in the Hermitian case. Conversely, the non-Hermitian protocol mitigates this degradation, i.e. $\partial \mathbb{E}[S]/\partial p(\lambda)\approx 1$, by effectively attenuating the incident photon flux down to a linear, non-saturating regime through postselection. In this high-flux regime, the non-Hermitian sensing protocol offers an apparent practical advantage, as illustrated numerically in Fig.\,\ref{fig_SAT}.


To gain deeper analytical insight into the numerical results presented above, we evaluate the statistical moments $\mathbb{E}[S]$ and $\mathrm{Var}[S]$ within a perturbative framework. By invoking the law of total probability, the unconditional distribution of the registered photoelectron counts $k_i$ is expressed as the mixture
\begin{equation}
    P(k_i)=\sum_{n_i}P(k_i| n_i)\,P(n_i).
\end{equation}
The corresponding first two moments of the distribution are given by
\begin{equation}
\begin{aligned}
    \mathbb{E}[k_i]
    &=
    \sum_{n_i}
    P(n_i)\,
    \mathbb{E}[k_i| n_i],
    \\
    \mathbb{E}[k_i^2]
    &=
    \sum_{n_i}
    P(n_i)\,
    \mathbb{E}[k_i^2| n_i].
\end{aligned}
\end{equation}
By utilizing the law of total variance, the total fluctuation profile of the output signal evaluates to
\begin{equation}
\begin{aligned}
   \mathrm{Var}(k_i)
    &=
    \mathbb{E}_{n_i}
    \!\left[
    \mathrm{Var}(k_i|n_i)
    \right]
    +
    \mathrm{Var}_{n_i}
    \!\left(
    \mathbb{E}[k_i|n_i]
    \right).
\end{aligned}
\end{equation}
Under a standard continuous approximation applied to the discrete Gaussian model, the conditional variance reduces to the constant electronic readout noise floor, $\mathrm{Var}(k_i|n_i)=\sigma_{\mathrm{r}}^2$. This simplifies the total channel variance to
\begin{equation}
    V_i\equiv\mathrm{Var}(k_i)
    =
    \sigma_{\mathrm{r}}^2
    +
    \mathrm{Var}_{n_i}
    \!\left[
    \bar{k}_i(n_i)
    \right],
\end{equation}
where the first term represents the intrinsic electronic noise background, and the second term encapsulates the propagation of shot-noise photon fluctuations through the nonlinear detector response.

Given that the incident photon distribution $P(n_i)$ is Poissonian with mean $\bar{n_i}$, we can evaluate these expectations exactly. Recalling the standard probability generating function for a Poisson random variable, $\mathbb{E}\!\left[a^{n_i}\right]=\exp\!\left[\bar{n}_i(a-1)\right]$, and identifying $a=e^{-1/n_{\mathrm{sat}}}$, the exact analytical expression for the mean output signal simplifies to
\begin{equation}
\label{Eq:ds_meanS}
    \mu_i
    \equiv
    \mathbb{E}[k_i]
    =
    k_{\mathrm{m}}
    \left[
    1-
    \exp\!\left(
    \bar{n}_i
    \left(e^{-1/n_{\mathrm{sat}}}-1\right)
    \right)
    \right].
\end{equation}
Following a parallel derivation for the second moment, the total output variance can be written in the compact closed form
\begin{equation}
\label{Eq:ds_varS}
\begin{aligned}
    V_i
    =
    \sigma_{\mathrm{r}}^2
    +
    k_{\mathrm{m}}^2
    \left[
    e^{\bar{n}_i\left(e^{-2/n_{\mathrm{sat}}}-1\right)} -
    e^{2\bar{n}_i\left(e^{-1/n_{\mathrm{sat}}}-1\right)}
    \right].
\end{aligned}
\end{equation}
Eqs.\,\eqref{Eq:ds_meanS} and \eqref{Eq:ds_varS} provide a fully self-contained description of the non-linear detector statistics. We now exploit these closed-form expressions to analyze how detector saturation modifies the normalized population estimator $S$. In the weak-saturation regime, where the mean incident photon numbers satisfy $1\ll \bar{n}_0,\bar{n}_1\ll n_{\mathrm{sat}}$, the exponential transfer function can be expanded perturbatively. To second order in the inverse saturation scale $1/n_{\mathrm{sat}}$, the mean output signal reduces to
\begin{equation}
\label{Eq:ds_channelmean}
    \mu_i
    \approx
    \frac{\bar{n}_i}{n_{\mathrm{sat}}}k_{\mathrm{m}}
    -
    \frac{\bar{n}_i+\bar{n}_i^2}{2n_{\mathrm{sat}}^2}
    k_{\mathrm{m}},
\end{equation}
whereas the leading-order correction to the channel variance yields
\begin{equation}
\label{Eq:ds_channelVar}
    V_i
    \approx
    \sigma_{\mathrm{r}}^2
    +\frac{\bar{n}_i}{n_{\mathrm{sat}}^2} k_{\mathrm{m}}^2.
\end{equation}
Under the assumption that the electronic readout noise is subordinate to the photon-induced fluctuations, i.e. $\sigma_{\mathrm{r}}^2 \lesssim \bar{n}_ik_{\mathrm{m}}^2/n_{\mathrm{sat}}^2$, the channel variance satisfies the inequality $V_i\ll \mu_i^2$. 
Crucially, Eq.\,\eqref{Eq:ds_channelmean} demonstrate that the conventional linear photon-counting scaling is recovered at leading order ($\sim 1/n_{\mathrm{sat}}$). The structural effects of detector saturation emerge systematically as higher-order non-linear corrections, shifting the effective response of the sensing system.

Under the physically reasonable assumption of uncorrelated channel statistics, we perform a small-fluctuation expansion by writing $k_i=\mu_i+\Delta k_i$, where $\mathbb{E}[\Delta k_i]=0$. A first-order multivariate Taylor expansion of $S$ in Eq.\,\eqref{Eq: electrProb} around the deterministic mean values yields the stochastic perturbation
\begin{equation}
    S
    \approx
    \frac{\mu_0}{\mu_0+\mu_1}
    +
    \frac{
    \mu_1\Delta k_0-\mu_0\Delta k_1
    }{
    (\mu_0+\mu_1)^2
    }.
\end{equation}
Taking the statistical expectation values, the ensemble average of the estimator reduces to the deterministic ratio
\begin{equation}
    \mathbb{E}[S_0]
    \approx
    \frac{\mu_0}{\mu_0+\mu_1},
\end{equation}
while its variance evaluates to the standard quadrature form
\begin{equation}
    \mathrm{Var}[S]
    \approx
    \frac{
    \mu_1^2V_0+\mu_0^2V_1
    }{
    (\mu_0+\mu_1)^4
    }.
\end{equation}
By further substituting the weak-saturation expansion in Eqs.\,\eqref{Eq:ds_channelmean} and \eqref{Eq:ds_channelVar} and keeping terms up to $\mathcal{O}(1/n_{\mathrm{sat}})$, the expectation of the population estimator evaluates to
\begin{equation}
\label{ds_approxSmean}
\begin{aligned}
    \mathbb{E}[S] &\approx
    \frac{\bar{n}_0}{\bar{n}_0+\bar{n}_1}
    +
    \frac{
    \bar{n}_0\bar{n}_1
    \left(
    \bar{n}_1-\bar{n}_0
    \right)
    }{
    2n_{\mathrm{sat}}
    \left(
    \bar{n}_0+\bar{n}_1
    \right)^2
    }\\
    &=
    p(\lambda)
    +
    \frac{n_{\mathrm{eff}}}{2n_{\mathrm{sat}}}
    p(\lambda)\left[1-p(\lambda)\right]\left[1-2p(\lambda)\right],
\end{aligned}
\end{equation}
where we have utilized the relations $\bar{n}_0=n_{\mathrm{eff}}p(\lambda)$ and $\bar{n}_1=n_{\mathrm{eff}}[1-p(\lambda)]$. Correspondingly, the leading-order weak-saturation approximation for the estimator variance yields
\begin{equation}
\begin{aligned}
    \mathrm{Var}[S]
    &\approx
    \frac{
    \bar{n}_0\bar{n}_1
    }{
    \left(
    \bar{n}_0+\bar{n}_1
    \right)^3
    }
    +
    \left(
    \frac{n_{\mathrm{sat}}}{k_{\mathrm{m}}}
    \right)^2
    \frac{
    \sigma_{\mathrm{r}}^2
    \left(
    \bar{n}_0^2+\bar{n}_1^2
    \right)
    }{
    \left(
    \bar{n}_0+\bar{n}_1
    \right)^4,
    }\\
    & \approx \frac{1}{n_{\mathrm{eff}}} p(\lambda)\left[1-p(\lambda)\right],
\end{aligned}
\end{equation}
where the final line holds provided the electronic readout noise is negligible compared to the photon-induced fluctuations, i.e. $\sigma_{\mathrm{r}}^2 \ll \bar{n}_ik_{\mathrm{m}}^2/n_{\mathrm{sat}}^2$. Under this condition, $\mathrm{Var}[S]$ remains inversely proportional to the effective photon resource, faithfully preserving the SQL scaling.

To investigate how the signal slope is modified by saturation, we take the partial derivative of Eq.\,\eqref{ds_approxSmean} with respect to the ideal projection probability,
\begin{equation}
    \frac{\partial \mathbb{E}[S]}{\partial p(\lambda)} = 1+\frac{n_{\mathrm{eff}}}{2n_{\mathrm{sat}}}\left[1-6p(\lambda)+6p(\lambda)^2\right].
\end{equation}
In the non-Hermitian protocol, the first-order correction term in the above equation is completely suppressed because the postselected resource is kept small ($n_{\mathrm{eff}} \ll n_{\mathrm{sat}}$). In contrast, this term cannot be neglected in the conventional Hermitian baseline, where the full input resource drives the system into the saturating regime ($n_{\mathrm{eff}} = n_{\mathrm{in}}\sim n_{\mathrm{sat}}$). For instance, at the typical operating point of Ramsey interferometry [$p(\lambda) = 1/2$], this saturation correction evaluates to $-n_{\mathrm{eff}}/4n_{\mathrm{sat}}$, reflecting the degradation of the measured slope in the Hermitian case.

Crucially, this perturbative analysis reveals that detector saturation does not inherently alter the fundamental SQL noise scaling law. Instead, it systematically degrades the overall sensitivity by suppressing the effective signal slope through the non-linear transfer function of the detection channels. In this regime, the unique capability of non-Hermitian sensing protocols to compress the identical amount of parameter information into a significantly smaller postselected ensemble endows them with a distinct practical advantage. By preventing the detectors from entering the sublinear saturation regime while preserving the information content, the non-Hermitian architecture bypasses the hardware-induced clipping that severely limits its conventional Hermitian counterpart.

\subsection{Systematic Parameter Offsets}
Beyond the detection noise during state readout process, the achievable precision is often fundamentally limited by our incomplete knowledge of the sensing apparatus itself. This epistemic uncertainty manifests as systematic parameter offsets that, unlike statistical fluctuations, cannot be suppressed simply by increasing the measurement repetitions $N$.

To formalize this framework, consider a scenario where the measurement outcome $\hat{p}$ depends not only on the target parameter $\lambda$, but also on a set of nuisance parameters governing the experimental setting. We categorize these into two distinct classes: fluctuating parameters $\hat{\kappa}$, and fixed unknown parameters $\zeta$. The fluctuating term $\hat{\kappa}$ varies stochastically between individual measurements (e.g., shot-to-shot laser intensity drift). For the $j$-th measurement, its local value $\hat{\kappa}_j$ is drawn from a distribution with mean $\bar{\kappa}$ and variance $(\Delta \kappa)^2$. In contrast, the parameter $\zeta$ represents a static quantity across the entire ensemble of $N$ repetitions but has a systematic calibration offset (e.g., a fixed misalignment or cavity detuning). Specifically, while its true physical value is denoted as $\zeta_r$, the observer lacks direct access to $\zeta_r$ and must rely instead on a prior estimation
$\zeta_p$ characterized by an uncertainty $(\Delta \zeta)^2$.

Assuming that the conditional probability density of the measurement outcomes is governed by $\hat{p}\sim P(\hat{p}|\lambda,\zeta,\hat{\kappa})$, the local expectation value of $\hat{p}$ for a given configuration  $\{\lambda,\zeta,\hat{\kappa}\}$ is expressed as 
\begin{equation}
    S(\lambda,\zeta,\hat{\kappa})\equiv \mathbb{E}[\hat{p}]=\int \hat{p} \cdot P(\hat{p}\mid\lambda,\zeta,\hat{\kappa})d\hat{p}.
\end{equation}
Let $\hat{p}_j$ denote the $j$-th individual measurement outcome sampled from the physical distribution $P(\hat{p}_j|\lambda_r,\zeta_r,\hat{\kappa}_j)$ where $\lambda_r$ represents the true value of the target parameter. The single-shot expectation value is then given by
\begin{equation}
    \mathbb{E}[\hat{p}_j] = S(\lambda_r,\zeta_r,\hat{\kappa}_j).
\end{equation}
To reconstruct the target parameter, the observer typically constructs an estimator $\hat{\lambda}$ by mapping the theoretical signal expectation evaluated at the nominal prior $\zeta_p$ to the empirical mean of the data ensemble,
\begin{equation} 
\label{Eq:lambda-estimator} 
S(\hat{\lambda},\zeta_p, \bar{\kappa})=\frac{1}{N}\sum_{j=1}^N \hat{p}_j.
\end{equation}
It is critical to note that the use of the prior estimate $\zeta_p$, rather than the true value $\zeta_r$, introduces a systematic discrepancy in the estimation process. To evaluate the resulting error, we expand Eq.\,\eqref{Eq:lambda-estimator} via a multivariate Taylor expansion around the actual operating point $\{\lambda_r,\zeta_r, \bar{\kappa}\}$ by subtracting the true ensemble mean $\sum_{j=1}^N S(\lambda_r,\zeta_r,\hat{\kappa}_j)/N$ from both sides, yielding
\begin{equation}
\begin{aligned}
  & \frac{1}{N}\sum_{i=1}^N (\hat{p}_i-\mathbb{E}[\hat{p}_i]) \\
  = &\frac{\partial S}{\partial \lambda}(\hat{\lambda}-\lambda_r)+\frac{\partial S}{\partial \zeta}(\zeta_p-\zeta_r)-\frac{1}{N}\sum_{i=1}^N \frac{\partial S}{\partial \kappa}(\kappa_i-\bar{\kappa}).
\end{aligned}
\end{equation}
From this linear relation, the parameter estimation variance, which quantifies the ultimate measurement precision of $\lambda$, can be formulated as
\begin{equation}
\begin{aligned}
  &(\delta\lambda)^2 = \mathbb{E}[(\hat{\lambda}-\lambda_r)^2]\\
  & = \frac{1}{|\partial_\lambda S|^2} \left[ \frac{(\Delta p)^2 + |\partial_\kappa S|^2(\Delta\kappa)^2}{N} + |\partial_\zeta S|^2(\Delta\zeta)^2 \right].
\end{aligned}
\end{equation}
Here, the three distinct terms represent the fundamental contributions to the error budget. The first term, characterized by the single-shot readout variance $(\Delta p)^2 = \sum_{i=1}^N \mathbb{E}[(\hat{p}_i-\mathbb{E}[\hat{p}_i])^2]/N$, captures the intrinsic quantum and classical readout noise. The second term arises from the stochastically fluctuating parameters, and both scale as $1/N$, following the standard quantum limit. Crucially, the third term represents the persistent influence of the systematic calibration error, which behaves as a $N$-independent precision floor. In the context of the noise decomposition framework introduced in Eq.\,\eqref{Eq:noiseDecomposition}, this systematic contribution corresponds to the non-standard scaling component,
\begin{equation}
    \sigma_{\mathrm{nSQL}}^2 = |\partial_\zeta S|^2(\Delta\zeta)^2.
\end{equation}
While conventional statistical averaging protocols can only minimize the first two terms, non-Hermitian sensing offers a unique advantage against this rigid systematic bottleneck. By engineering the system to operate in a regime of divergent susceptibility where $|\partial_\lambda S|\sim \chi_{\mathrm{nH}}\gg 1$, the parameter-induced signal is strongly amplified. This amplification effectively suppresses the impact of both the fluctuating technical noise and the persistent calibration floor $\sigma_{\mathrm{nSQL}}^2$, unlocking high-precision regimes that are otherwise inaccessible to conventional sensors bounded by systematic errors.

\section{Information-theoretic analysis}
\label{sec:information}
In this section, we quantify the metrological advantage of non-Hermitian protocols within the information-theoretic framework of parameter estimation theory. Our analysis focuses on two fundamental figures of merit: the classical Fisher information (FI) $F_\lambda$ and the quantum Fisher information (QFI), $\mathcal{F}_\lambda$ ~\cite{PhysRevLett.72.3439}. These quantities establish the ultimate precision limits for estimating an unknown parameter $\lambda$ via the Cramér–Rao bound:
\begin{equation} 
(\delta \lambda)^2 \geq \frac{1}{N F_\lambda} \geq \frac{1}{N \mathcal{F}_\lambda}, 
\end{equation}
where $N$ denotes the number of independent measurement repetitions. Therefore, a larger QFI or FI signifies a better achievable estimation uncertainty.

\subsection{Basic Formulas}
The QFI represents the maximum extractable information encoded in a parameter-dependent quantum state $\rho_\lambda$, optimized over all physically realizable measurement strategies. It is formally defined as
\begin{equation} 
\mathcal{F}_\lambda = \mathrm{Tr}\left[\rho_\lambda L_\lambda^2\right], 
\end{equation}
where the symmetric logarithmic derivative $L_\lambda$ is the Hermitian operator satisfying the Lyapunov equation 
\begin{equation}
    \frac{\partial\rho_\lambda}{\partial \lambda} = \frac{1}{2}\left(\rho_\lambda L_\lambda+L_\lambda\rho_\lambda\right).
\end{equation}
For pure-state metrology, where $\rho_\lambda = |\psi_\lambda\rangle \langle \psi_\lambda|$, the QFI (by solving the above two equations) simplifies to the following form,
\begin{equation} 
\mathcal{F}_\lambda = 4\left(\langle \partial_\lambda \psi_\lambda | \partial_\lambda \psi_\lambda\rangle - |\langle \partial_\lambda \psi_\lambda | \psi_\lambda\rangle|^2 \right).
\label{Eq:QFI} 
\end{equation}
While the QFI sets the fundamental quantum ceiling, the classical FI, $F_\lambda$, quantifies the information actually retrieved by a specific measurement apparatus. Given a measurement outcome $x$, governed by the conditional probability distribution $P(x|\lambda)$, the FI is defined as the variance of the score function,
\begin{equation}
    F_\lambda = 
    \int \mathrm{d}x P(x|\lambda)\left[\frac{\partial}{\partial \lambda}\ln P(x|\lambda)\right]^2.
\end{equation}
Physically, $F_\lambda$ characterizes the sensitivity of the log-likelihood function $\ln P(x|\lambda)$ to infinitesimal changes in $\lambda$. Broad, overlapping distributions yield a diminished FI, whereas highly localized distributions with strong parametric dependence result in a large FI value. In experimental contexts, $F_\lambda$ serves as the primary practical benchmark, capturing the delicate interplay between the intrinsic quantum sensitivity and the inevitable resolution degradation induced by readout noise and finite sampling statistics.

\begin{figure}[t]
    \centering
    \includegraphics[width=0.95\linewidth]{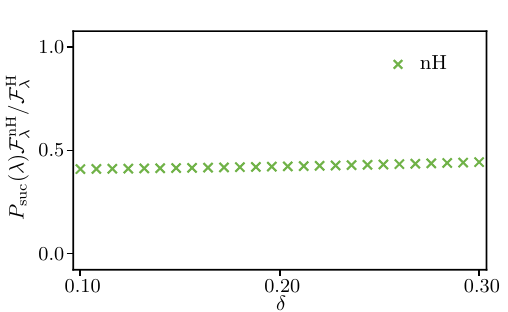}
    \caption{{\bf  Absence of QFI advantage under postselection.} 
We plot the operational QFI, $P_{\mathrm{suc}}(\lambda)\mathcal{F}_\lambda^{\mathrm{nH}}$, of the non-Hermitian protocol normalized to the Hermitian baseline ($F_\lambda^{\rm H}=4t^2$) as a function of the control parameter $\delta$. For each $\delta$, the non-Hermitian QFI is maximized over the parameter $\lambda$. When accounting for the success probability $P_{\mathrm{suc}}(\lambda)$, the operational QFI remains strictly below the optimized Hermitian baseline, confirming that the postselected non-Hermitian protocol provides no intrinsic metrological advantage in the ideal setting. Parameters of non-Hermitian sensing based on the dilation framework in Eq.\,\eqref{Eq:DilationHtot} are set to $a=1$, $\mathcal{E}=0.5$ and $t=\pi/(2\mathcal{E})$.}
    \label{fig_QFI}
\end{figure}

\subsection{Operational QFI in Hermitian Dilation Framework}
The apparent metrological advantage of non-Hermitian systems frequently manifests as a spurious divergence in the QFI when it is calculated directly from the renormalized version of the non-Hermitian state $|\psi(t)\rangle$ in Eq.\,\eqref{Eq:nonHstate}. A rigorous metrological assessment requires a faithful accounting of the physical protocols implementing such dynamics. In the Naimark dilation protocol in Eq.\,\eqref{Eq:DilatedState} to achieve non-Hermitian sensing, while the QFI evaluated using the renormalized state from $\ket{\psi (\lambda, t)}$ yields a spurious divergence, i.e.   $\mathcal{F}_\lambda^{\mathrm{nH}} \to \infty $ as $\delta\to 0$, a physically meaningful operational QFI must be weighted by the postselection probability i.e. $P_{\mathrm{suc}}(\lambda) \mathcal{F}_\lambda^{\mathrm{nH}}$. This correction recovers the finite precision limits illustrated in Fig.\,\ref{fig_QFI}. Under this rigorous formulation, the operational QFI of the non-Hermitian configuration never exceeds the baseline imposed by the optimal Hermitian sensing protocol (i.e. Ramsey interferometry).

\subsection{FI Accounting for Technical Noises}
While the QFI establishes the ultimate theoretical ceiling for sensing precision, its operational utility in non-Hermitian systems is fundamentally limited by its failure to account for classical technical noise. To bridge the gap, we evaluate the classical FI. Unlike the QFI, the FI directly characterizes the statistics of physical measurement outcomes, providing a natural mathematical framework to incorporate the non-reducible technical noise—such as correlated background detection noise as well as the saturation of photon detector. 

In realistic sensing scenarios, deriving analytical closed-form expressions for the Fisher information under full noise models is often unfeasible. To evaluate the performance under realistic conditions, we numerically calculate the noisy probability distributions and use a finite-difference approximation to obtain the Fisher information. Specifically, we start from the theoretical projection probability onto the state $|0\rangle$. Experimental imperfections are then incorporated by generating Monte Carlo measurement samples governed by the noise models described in the previous sections. Finally, the empirical Fisher information is extracted from the parametric derivative of these simulated noisy distributions.

To incorporate the effects of background detection noise, the photon counts from the quantum sensor during the $j$-th measurement trials, $x_j$, are sampled according to Eq.\,\eqref{Eq:phtonGaussian}. The total detected signal $X_j$ is modeled as a combination of these sensor counts and the correlated background photons $\xi_j$. We simulate the background fluctuations $\xi_j$  using the correlated multivariate Gaussian noise in Eq.\eqref{Eq: cbn_xi}.
The resulting probability distribution of the estimator $\hat{p}$ in Eq.\,\eqref{Eq: bdn_estimator}, denoted by $P(\hat{p}|\lambda)$, is subsequently reconstructed via a large ensemble of numerical Monte Carlo samples. Crucially, to ensure a rigorous and equitable comparison between Hermitian and non-Hermitian sensing protocols, the effective number of measurement repetitions utilized in Eq.\,\eqref{Eq: bdn_estimator} must be constrained. Specifically, for a total budget of available experimental trials $N=N_{\mathrm{t}}$, we allocate the full resource $N=N_{\mathrm{t}}$ to the conventional Hermitian baseline, whereas the resource allocated to the non-Hermitian protocol is penalized by its postselection success probability, yielding $N=P_{\mathrm{suc}}(\lambda) N_{\mathrm{t}}$.

\begin{figure}[t]
    \centering
    \includegraphics[width=0.95\linewidth]{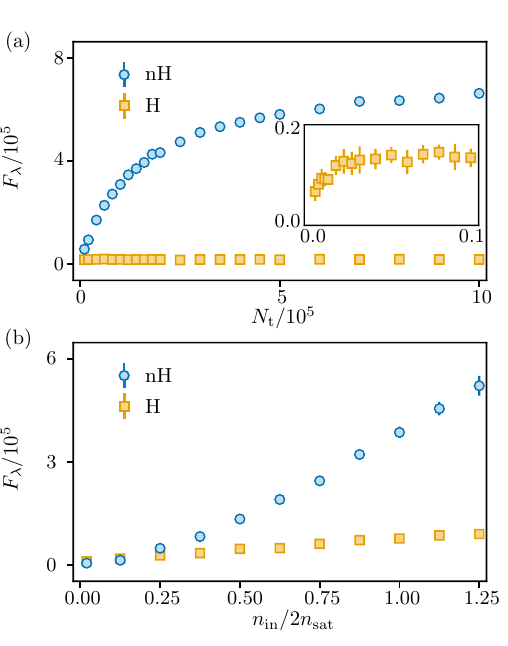}
    \caption{{\bf Fisher information in the presence of technical noises.}
  (a) FI under fully correlated background noise, plotted as a function of the number of repeated measurements $N_{\mathrm{t}}$. 
  (b) FI under detector saturation noise as a function of the normalized input photon budget $n_{\rm in}/(2n_{\rm sat})$. 
  In both panels, the postselection success probability is incorporated into the effective non-Hermitian sample size, and error bars indicate the standard deviation over ten independent Monte Carlo realizations. These results demonstrate that the non-Hermitian protocol can exhibit a practical FI advantage in the presence of realistic detection noise.
  For the numerical simulations, the common parameters for the non-Hermitian protocol across both panels are fixed at $a=1$, $\mathcal{E}=0.5$ and $t=\pi/(2\mathcal{E})$. In (a), the non-Hermitian protocol operates with $\delta=0.225$ and $\lambda=-0.1279$ with  $\Delta\lambda=1.5\times10^{-3}$, under readout noise parameters $\sigma_\xi=0.06$, $x_{\left | 0 \right \rangle}=7$, $x_{\left | 1 \right \rangle}=5$ and $\sigma=1$. In (b), the non-Hermitian parameters are set to $\delta=0.12$, $\lambda_{\mathrm{}}=-0.005$ and $\Delta\lambda=10^{-3}$ with electronic readout noise parameters given by $n_{\mathrm{sat}}=1000$, $k_{\mathrm{m}}=4000$ and $\sigma_{\mathrm{r}}=10$.}
    \label{fig_FI}
\end{figure}

To model the saturation effects of the photon detectors, we denote the raw photon counts arriving at the two spatial detection channels as $n_0$ and $n_1$. These counts follow independent Poisson distributions governed by Eq.\,\eqref{Eq: ds_photons}. Consequently, the non-linear photoelectron output of detection channel $\mu \in \{0, 1\}$ is modeled via the standard saturation characteristic
\begin{equation}
k_{\mu}
=
k_{\max}
\left[
1-\exp\left(
-\frac{n_{\mu}}{n_{\mathrm{sat}}}
\right)
\right] +\epsilon_{\mu},
\end{equation}
where $\epsilon_{\mu}\sim\mathcal N(0,\sigma_{\mathrm r}^2)$ accounts for independent Gaussian readout electronic noise. Similarly, the operational probability distribution $P(S|\lambda)$ of the estimator defined in Eq.\,\eqref{Eq: electrProb} is reconstructed empirically through a large ensemble of independent Monte Carlo simulation trials.

To evaluate the Fisher information from the simulated data, we employ a histogram binning technique. Let $p_m(\lambda)$ denote the integrated probability of $\hat{p}$ in Eq.\,\eqref{Eq: bdn_estimator} or $S$ in Eq.\,\eqref{Eq: electrProb} in the above cases falling within the $m$-th bin. The parametric derivative with respect to
$\lambda$ is evaluated via a finite-difference approximation,
\begin{equation}
\partial_\lambda p_m(\lambda)
\approx
\frac{p_m(\lambda+\Delta\lambda)-p_m(\lambda)}{\Delta\lambda}.
\end{equation}
The operational Fisher information is subsequently computed via the discrete summation
\begin{equation}
F_\lambda
\approx
\sum_m \frac{1}{\bar p_m(\lambda)}
\frac{\big[p_m(\lambda+\Delta\lambda)-p_m(\lambda)\big]^2}
{\Delta\lambda^2},
\end{equation}
where $\bar p_m(\lambda)$ implements a symmetric regularization scheme in the denominator to ensure numerical stability,
\begin{equation}
\bar{p}_m (\lambda)
=
\frac{
\hat{p}_m(\lambda+\Delta\lambda)
+
\hat{p}_m(\lambda)
}{2}
+
\varepsilon.
\end{equation}
Here, $\varepsilon=10^{-12}$ is a small numerical regularizer introduced to suppress artificial divergences arising from low-count or empty bins. Finally, the statistical uncertainty and confidence intervals of the numerically extracted $F_\lambda$ are quantified by bootstrapping the estimation procedure over 10 independent ensemble realizations.

Owing to the significantly enhanced parametric susceptibility ($\chi_{\mathrm{nH}}\gg 1$) provided by the non-Hermitian protocol relative to the non-reducible technical noise floor $\sigma_{\mathrm{nSQL}}$, the FI displayed in Fig.\,\ref{fig_FI} exhibits a substantial enhancement over the conventional Hermitian baseline. This behavior establishes a clear operational advantage for non-Hermitian sensing configurations when subjected to realistic experimental constraints from an information-theoretic perspective.

\section{Conclusion and Outlook}
\label{sec:discussion}
In conclusion, we have provided a comprehensive analysis of non-Hermitian sensing, focusing on the transition from fundamental quantum limits to practical metrological performance. While recent theoretical findings have underscored that non-Hermitian protocols offer no intrinsic advantage over Hermitian counterparts in the ideal SQL regime—once the physical overhead of non-unitary evolution is accounted for—our work demonstrates that this ``no-go'' result does not extend to realistic experimental settings. Under various technical noise models, we have shown that non-Hermitian platforms can yield significant precision enhancements in the presence of instrumental and environmental imperfections.

The physical origin of this resilience lies in the significantly amplified susceptibility inherent to non-Hermitian protocols. This mechanism boosts the signal response, allowing it to effectively overcome the constant floor of classical technical noise, which often defines the operational ceiling in the laboratory. Our results thus reconcile the perceived lack of a universal quantum advantage with the diverse experimental reports of enhanced sensitivity, clarifying that the true value of non-Hermiticity lies in its robustness against practical constraints rather than its scaling in the idealized limit.

Looking forward, our findings open several avenues for future research. Experimentally, the identified noise-resilient regimes provide a concrete roadmap for implementing high-precision sensors in platforms ranging from solid-state spins to superconducting circuits, where technical noise remains a primary bottleneck. Theoretically, extending our framework to multi-parameter estimation and non-Hermitian many-body systems could reveal even more sophisticated strategies for noise suppression. By shifting the focus toward the functional utility of non-Hermitian physics in realistic environments, this work paves the way for a new generation of sensors that are not only theoretically intriguing but also practically robust in real-world applications.

\section{Acknowledgments}
This work was supported by the National Natural Science Foundation of China (12304572, 12425414, 12304571, U25D9006), and Quantum Science and Technology-National Science and Technology Major Project (2024ZD0300900,2024ZD0300902). Y.C. also acknowledges the support of the Fundamental Research Support Program of Huazhong University of Science and Technology (2025BRB001) and the Major Science and Technology Project of Hubei Province (2025BEA001). The authors declare no competing financial or non-financial interests.

\newpage
\bibliography{reference}

\end{document}